# Paths to Testing: Why Women Enter and Remain in Software Testing?

Kleice Silva
CESAR School
Recife, PE, Brazil
kmbs@cesar.school

Ann Barcomb
University of Calgary
Calgary, AB, Canada
ann@barcomb.org

Ronnie de Souza Santos
University of Calgary
Calgary, AB, Canada
ronnie.desouzasantos@ucalgary.com

## ABSTRACT

**Background**. Women bring unique problem-solving skills to software development, often favoring a holistic approach and attention to detail. In software testing, precision and attention to detail are essential as professionals explore system functionalities to identify defects. Recognizing the alignment between these skills and women's strengths can derive strategies for enhancing diversity in software engineering. **Goal**. This study investigates the motivations behind women choosing careers in software testing, aiming to provide insights into their reasons for entering and remaining in the field. **Method**. This study used a cross-sectional survey methodology following established software engineering guidelines, collecting data from women in software testing to explore their motivations, experiences, and perspectives. **Findings**. The findings reveal that women enter software testing due to increased entry-level job opportunities, work-life balance, and even fewer gender stereotypes. Their motivations to stay include the impact of delivering high-quality software, continuous learning opportunities, and the challenges the activities bring to them. However, inclusiveness and career development in the field need improvement for sustained diversity. **Conclusion**. Preliminary yet significant, these findings offer interesting insights for researchers and practitioners towards the understanding of women's diverse motivations in software testing and how this understanding is important for fostering professional growth and creating a more inclusive and equitable industry landscape.

## 1 INTRODUCTION

Despite the historical underrepresentation and gender-related challenges in the technology field, women have made remarkable contributions to software engineering [1, 9]. Their impact on the area is far-reaching, spanning from pioneering coding techniques to shaping cutting-edge software applications [9]. Over the years, women have demonstrated their expertise in diverse aspects of software development, including software design, programming, quality assurance, and beyond, which has been essential in driving innovation and fostering the development of more inclusive and user-friendly software applications [5, 28].

Previous studies have highlighted the unique set of problem-solving skills that women bring to work [15, 25]. While individual differences exist, some common trends have emerged. In problem-solving tasks, like software development, men tend to lean towards quicker decision-making and assertive actions when faced with challenges [11, 18, 29]. In contrast, women often adopt a more holistic approach, considering various aspects of a problem. Women's attention to detail, thoroughness, and capacity to consider multiple perspectives contribute to their effectiveness in tasks that demand precision [11, 18, 29].

In software development, software testing is a critical phase of the process where professionals are tasked with applying a specific skill set to identify faults [8]. Typically, testing professionals demonstrate remarkable attention to detail and explore the system functionalities to detect defects, considering various perspectives, user scenarios, and system configurations [4]. In this context, precision is paramount for testing professionals to identify bugs and deliver high-quality applications.

Looking at the literature, we can observe that many skills essential for excelling in software testing [4] align with qualities associated with how women often approach their work [11, 18, 29]. These qualities include being more other-oriented, exercising greater caution, and displaying heightened sensitivity to varying conditions and factors [18]. This observation triggered the motivation for this study: to explore the experience of women in software testing, with a specific focus on an industrial perspective.

The main contribution of this research is the discussion of the perceptions of software professionals who identify as women and work in the industrial context of software testing, emphasizing the distinctive viewpoints they bring to the field. Our findings contribute to a more inclusive and diversified software testing landscape by revealing motivations, challenges, and opportunities for future investigations regarding the role of individual characteristics (e.g., gender) in software engineering. Hence, to address this topic, we focused on the following research question: ***RQ*** *What are the primary reasons that drive women to choose software testing as their career path, and what motivates them to stay?*

Building upon this introduction, our study is structured as follows: In Section 2, we provide a detailed overview of our research methodology. Section 3 presents our findings. Section 4 is centered on discussions and the implications drawn from our study. Section 5 discussed threats to validity. Finally, Section 6 focuses on our research contributions and offers a summary of our work.

## 2 METHOD

To investigate the experiences of women in the software testing field, we opted for a real-world, industry-focused approach based on a cross-sectional survey [10, 16]. Industrial surveys play a crucial role in software engineering, offering valuable insights into various aspects of software development from the viewpoint of those actively engaged in the field [27].

Despite occasional shortcomings such as a lack of novelty, limited geographical coverage, and sampling constraints, these studies continue to be indispensable [27] as their findings not only contribute to improving discussions on existing evidence but also serve as a foundational basis for future, more in-depth investigations on a given topic. As an example, several industrial surveys on software testing can be identified in the literature, including [6, 12, 14, 22, 23]

Hence, aligned with well-established software engineering research guidelines [19, 20], we sought insights from women in the field of software testing. This involved exploring their career interests, motivations for choosing this path, and their experiences within the industry. The step-by-step process of how we designed and conducted our survey is presented below.

### 2.1 Questionnaire

Questionnaires are the primary instrument used in surveys, particularly when collecting data from practitioners across diverse fields, including software engineering [10, 16]. They offer a structured and standardized approach that ensures each participant is exposed to an identical set of questions. This standardization significantly enhances the reliability and comparability of responses.

In this research, our questionnaire was thoughtfully structured to encompass a combination of multiple-choice and open-ended questions, offering women engaged in software testing the opportunity to provide insights into their experiences. We included a series of queries focused on participants' professional and personal backgrounds, motivations, and expectations, thereby enhancing the richness of the data collected in the study.

Following the formulation of the questions, a pilot questionnaire underwent validation by two female testing professionals who were not included in the study sample. They offered insightful feedback, which led to refinements in sentence wording and the incorporation of a set of closed-ended questions related to intrinsic and extrinsic work motivations. The final version of the questionnaire is provided in Table 1.

### 2.2 Data Collection

This study employed convenience sampling [3], where participants were selected based on availability rather than through a random process. In studies that aim to gather preliminary insights, explore new research areas, or conduct pilot investigations, convenience sampling can provide a quick and practical means of data collection [2]. Hence, convenience sampling aligns with our specific goals, focusing on gathering initial insights into women's experiences in software testing and, based on the outcomes, preparing for a more comprehensive investigation by extending this survey using other sampling techniques or undertaking a qualitative approach in the future.

**Table 1: Survey Questionnaire**

---

1. You are invited to participate in our research designed to explore the motivations driving individuals who identify with the female gender to choose and pursue a career in software testing. If you choose to participate, rest assured that all gathered information will be handled exclusively for the purposes of this research. Your participation will be entirely anonymous, ensuring that you will not be identifiable in any publication related to the discussed topic. The estimated time to complete the survey is approximately 5 to 10 minutes. Your insights are invaluable and greatly appreciated. Do you agree to participate?
( ) Yes

---

2. How old are you?

3. What is your highest educational level?
( ) High-School
( ) Bachelor's degree
( ) Post-baccalaureate or Diploma
( ) Master's degree
( ) Ph.D.

4. How long have you been working with software testing?

5. What is your job level?
( ) Trainee
( ) Beginner
( ) Mid-level
( ) Senior
( ) Principal

6. Do you have any testing certification?

7. Do you have any other technical certification?

---

8. What are your main activities working in software testing?

9. Briefly comment on your journey into the QA field.

10. What motivations influenced your decision to pursue a career in this area?

11. Throughout your educational journey, was there a particular influence that guided you toward software testing? Feel free to share your experiences!

12. What aspects of software testing are currently exciting you the most?

13. Share with me what brings you fulfillment and happiness in your work within this area.

---

14. To what extent do you agree with the statements below?
A. I have the autonomy necessary to make the required changes in my tasks.
( ) Strongly Agree
( ) Agree
( ) Disagree
( ) Strongly Disagree

B. I am acknowledged for my work.
( ) Strongly Agree
( ) Agree
( ) Disagree
( ) Strongly Disagree

C. My voice and opinions as a testing professional are underestimated or ignored.
( ) Strongly Agree
( ) Agree
( ) Disagree
( ) Strongly Disagree

C. The prospect of promotions and advancements in my career is an important factor for me to continue in software testing.
( ) Strongly Agree
( ) Agree
( ) Disagree
( ) Strongly Disagree

---

Specifically, data collection for this study involved reaching out to participants through our extensive network of software professionals. Leveraging industry contacts, we distributed our questionnaire to key professionals in software companies and promoted the study on LinkedIn. In an unforeseen but valuable turn, some responses were obtained as participants shared the questionnaire with others within the targeted population, resembling a non-planned snowballing sampling approach. This recruitment strategy supported the identification of a diverse pool of participants and contributed to the richness of perspectives gathered in our study.

## 2.3 Data Analysis

Following the data collection phase, we applied the principles of descriptive statistics [13] to systematically investigate and summarize distribution patterns and the frequency of participants' responses. This method provided an insightful perspective on the quantitative data, enhancing our understanding of key aspects and trends within the sample.

In addition, we employed thematic analysis techniques [7] to delve into the detailed narratives provided by participants in response to open-ended questions. This analytical method enabled the extraction and identification of recurring themes and patterns within the qualitative data, providing an understanding of women's experiences and perspectives on software testing. Figure 1 illustrates our qualitative analysis strategy based on coding techniques and thematic analysis.

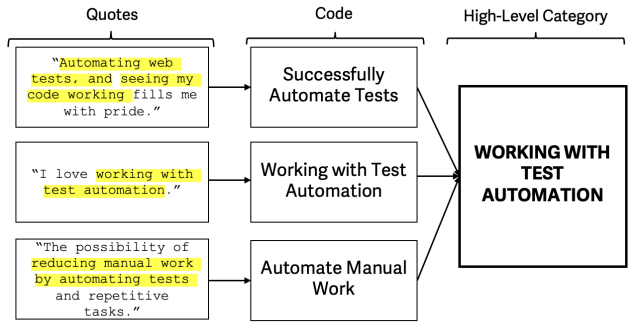

**Figure 1: Coding Process**

## 2.4 Ethics

In line with ethical standards, this study maintained participants' privacy by avoiding collecting personally sensitive information, including names, emails, or employer details, thus safeguarding their anonymity. All participants were explicitly informed about the scientific use of their data and willingly agreed to participate. For those expressing interest in participating in subsequent phases of our project, we invited them to contact us for potential interviews, assuring them of continued anonymity throughout any future stages of our research.

## 3 FINDINGS

Our findings encapsulate the experiences of 49 women occupying specialized roles in software testing, ranging from test engineers and test analysts to testing managers and quality assurance professionals (QAs). Our sample is composed of experienced professionals, with 39% of participants having more than five years of industrial expertise—this proportion aligns with the number of participants in senior or principal positions. Additionally, 33% of the sample holds a testing certification from the International Software Testing Qualifications Board, while 57% possess various other certifications, including those related to agile methodologies. An overview of our sample characteristics is presented in Table 2.

**Table 2: Demographics**

| Participants Profile | | | |
|---|---|---|---|
| Age | Up to 25 years old | 2 | (4.1%) |
| | 25-34 years old | 27 | (55.1%) |
| | 35-44 years old | 17 | (34.7%) |
| | 45+ years old | 3 | (6.1%) |
| Educational Level | High-School | 4 | (8.2%) |
| | Bachelor's degree | 15 | (30.6%) |
| | Post-baccalaureate | 29 | (59.2%) |
| | Master's degree | 1 | (2.0%) |
| Job Level | Trainee | 2 | (4.1%) |
| | Beginner | 17 | (34.7%) |
| | Mid-level | 12 | (22.5%) |
| | Senior | 15 | (30.6%) |
| | Principal | 3 | (6.1%) |
| Experience | Less than 1 year | 11 | (22.4%) |
| | 1-2 Years | 19 | (38.8%) |
| | 3-4 Years | 0 | (0%) |
| | 5+ Years | 19 | (38.8%) |
| Testing Certification | No | 33 | (67.3%) |
| | Yes | 16 | (32.7%) |
| Other Certification | No | 21 | (42.9%) |
| | Yes | 28 | (57.1%) |
| Working Environment | Office | 4 | (8.2%) |
| | Hybrid | 7 | (14.3%) |
| | Remote | 38 | (77.6%) |

### 3.1 Women Joining Software Testing: Motivations and Reasons

We identified at least seven factors that influence women to pursue a software testing career. These pivotal factors are presented in Table 3, with quotations directly extracted from participants' questionnaires. These excerpts capture the nuanced perspectives and diverse experiences of the women who participated in our study.

Twelve participants emphasized that software testing, as a facet of software development, provides ample opportunities, particularly for those entering the software industry. In contrast, nine participants discovered software testing through exploration and experimentation with various roles, eventually developing a passion for testing. Conversely, seven participants expressed that their skills shaped their inclination toward software testing—choosing this path because they possessed or cultivated the necessary knowledge and skills.

Among our participants, six individuals shared that software testing was not initially their primary choice – Organizational and managerial decisions shifted them from programming-focused roles, such as back-end development, into software testing. Additionally, five participants highlighted that recommendations from other professionals, particularly women in the field, played a crucial role in guiding them toward a career in software testing. Only four participants credited their university experiences, particularly through an undergraduate course focused on software testing, as a motivation to enter the field. Lastly, four participants emphasized that their decision was influenced by the potential to balance their career with personal tasks, particularly in terms of being close to or taking care of children.

While these women entered the field of software testing for diverse reasons, a prevailing consensus among the majority is their genuine enjoyment of quality-centric tasks. Some participants highlighted salary as a factor contributing to their contentment in working within the field. Interestingly, some professionals admitted that although testing was not their initial choice, their evolving experience led them to develop a genuine liking for software quality work. As a result, they express an intention to continue their careers in this domain.

**Table 3: Reasons to Join Software Testing**

| Factor | Quotations |
| --- | --- |
| Job Opportunities | - *"During my career transition, this field proved to be the most accessible entry point into the software industry for me."* (P08) <br> - *"The QA field offers many opportunities for beginners to start their careers."* (P10) <br> - *"It was the first opportunity I had as an intern, and resonating with the field, I chose to specialize in it."* (P24) |
| Exploring Careers | - *"I was assigned to an Android team, but I didn't enjoy working with code and programming, so I requested a relocation to another team, and they reassigned me as a manual tester."* (P17) <br> - *"I've worked as a screening and overflow operator, promoter, bartender, administrative assistant, and biologist, but none of these roles make me happy. Then, I tried the quality assurance field and found my true calling!"* (P27) <br> - *"By chance, I was a programmer and the testing opportunity came my way."* (P29) |
| Testing Skills | - *"Because I am highly curious and analytical."* (P15) <br> - *"I identified with the field because it aligns closely with my professional profile, which always emphasizes ensuring quality in everything I do."* (P37) <br> - *"I was already working in support for a sales system when I began identifying some bugs and became interested in software testing."* (P42) |
| Recommendations | - *"My Undergraduate Thesis advisor recommended me for a position in the field, and after successfully securing the role, I had no desire to transition to something else."* (P03) <br> - *"I entered the QA field based on the recommendation of a friend who was already working as a QA."* (P05) <br> - *"I started studying the field through a friend who worked in testing. She told me about it and recommended a course."* (P32) |
| University Experience | - *"Since my college days, I've been interested in the software quality field, as I realized that it wasn't confined to the stereotypical image associated with software professionals."* (P01) <br> - *"I became interested in the field through an elective course on Software Testing in the university."* (P02) <br> - *"I had a class on software quality at the university, and it turned into a habit to search for bugs on famous websites."* (P46) |
| Career Balance | - *"The possibility of working remotely, allowing me to take care of my child."* (P12) <br> - *"The opportunity to work from home, given the routine with my children, makes it challenging for me to work outside the home."* (P19) <br> - *"to try to find a remote work because I no longer had help to take care of my children."* (P33) |

### 3.2 The role of university bringing women to Software Testing

Even though some participants mentioned that their academic experiences helped them envision a career in software testing (Section 3.1), our questionnaire explicitly explored this aspect. However, consistent with earlier findings (Section 3.1), only a minority of women had any exposure to software testing during their academic years. Within the sample, seven participants reported some influence from their university experiences. Three of them highlighted the impact of a specific course on software testing in guiding their career choice. Two participants attributed their inspiration to teachers with industry testing experience who discussed the profession in class. Another participant emphasized that an assignment involving testing significantly influenced her decision. Finally, one participant cited a study group focused on testing certifications as her primary academic contact with the field.

### 3.3 Current Motivations to Remain in Software Testing

As detailed in Section 3.1, not all women in our sample initially intended to pursue careers in software testing. In fact, most found themselves in this field by chance—seizing opportunities, exploring scenarios, or being assigned to the role. Despite the unplanned nature of their entry, these women now express satisfaction with their careers in software testing. Over time, the work has become personally fulfilling, and various motivating factors drive their commitment to this field. These factors are illustrated with quotations in Table 4.

A significant majority of participants, more than half, indicated that their motivation comes from the meaningful role that software testing plays in delivering high-quality systems to society. Knowing that their work contributes to something with a positive impact on people's lives serves as a powerful motivator for them to stay in the field. Additionally, seven participants emphasized the constant exposure to learning opportunities—be it acquiring new knowledge or sharing their expertise with others—this dynamic learning context contributes to a positive perception of the field for them.

Moreover, seven participants emphasized that their motivation comes from engaging in coding tasks, particularly programming, as they are consistently involved in the testing automation process. This sentiment is further supported by three other women who perceive software testing as a challenging and dynamic field, which serves as a motivational factor. Lastly, a few participants in the sample expressed motivation through the excitement of participating in team activities (i.e., belonging to a team) or the opportunity to work remotely.

### 3.4 Women in the Team

As part of our investigation, we also explored several facets of team dynamics, aiming to understand how women perceive themselves and their roles as testing professionals within their teams. In this sense, participants provided insights into autonomy, the influence of their decisions, feedback mechanisms, and their perspectives on promotion and career development.

The results obtained for these factors are as follows: *Autonomy*: 24% of our sample indicated that they do not possess the level of autonomy they desire to carry out their work effectively; *Professional Voice*: 27% of participants expressed that they feel unheard when decisions come into play. These women reported experiencing underestimation or being overlooked in their work, leading to an increased sense of frustration; *Acknowledgement*: 94% of our sample expressed that receiving positive feedback about their work and projects is a significant source of motivation for them; *Promotion and Career Development*: 92% of these women perceive significant potential for promotions and career development in the field of software testing. This perception serves as a crucial factor driving their motivation to remain in the field.

**Table 4: Reasons to Stay in Software Testing**

| Factor | Quotations |
|---|---|
| Meaningful Work | - *"What excites me is knowing that I played a crucial role in the production process, ensuring and upholding its quality."* (P12)<br>- *"What truly brings me joy is when an entire system goes live, and people start using it. (...) knowing that I played a part in ensuring the quality and smooth operation."* (P40)<br>- *"Being in a project where different types of users can use something I helped build is truly fulfilling."* (P02) |
| Learning Opportunities | - *"The fact that I am always learning, and there are always new scenarios and realities, keeps me engaged."* (P01)<br>- *"As a team leader today, what motivates me is assisting people at various levels and being able to share my knowledge not only to the testing people but also in other areas."* (P38)<br>- *"It's the processes and the constant cycle of learning and teaching something to someone that motivates me."* (P36) |
| Working with Test Automation | - *"I love working with test automation."* (P03)<br>- *"The opportunity to reduce manual work by automating tests and repetitive tasks excites me."* (P31)<br>- *"Developing code brings me immense joy. I love seeing it in action, automating websites, and feeling proud as my code works independently."* (P19) |
| Challenging Tasks | - *"MOTIVATION lies in the CHALLENGES and opportunities that this dynamic field has to offer."* (P41)<br>- *"Learning from challenges."* (P47)<br>- *"Searching and finding what is not obvious [is challenging]"* (P10) |
| Possibility of Remote Work | - *"The possibility of working remotely, allowing me to take care of my child."* (P12)<br>- *"The possibility of working from home."* (P29)<br>- *"Salary and remote work opportunities."* (P39) |
| Engage in Teamwork | - *"Daily experience, teamwork, and processes."* (P18)<br>- *"I enjoy the community itself, where I've connected with fantastic individuals who motivate me to contribute even more to the field."* (P23) |

Our sample reports general positive experiences and perceptions within their teams. However, the findings indicate that there is room for improvement in terms of autonomy and the inclusion of women in team decisions, such as increasing the visibility of their contributions and opinions.

## 4 DISCUSSION

Our study provides an initial analysis of the motivations that drive women to pursue careers in software testing. The preliminary findings encourage further exploration of the subject, prompting us to develop strategies aimed at increasing the visibility of this field among women. The issue of invisibility in software testing, a topic discussed in prior research [8], underscores the potential repercussions of not having a motivated and skilled workforce in this crucial area of the software industry. Our study has yielded initial insights, indicating that women may be particularly attracted to the field for various reasons, including their skills and the fulfillment derived from impacting the delivery of high-quality software. Additionally, as we navigated through the evidence collected from our participants, we observed some interesting factors that deserve attention and might be important to explore.

Firstly, previous studies on remote and hybrid work in software engineering have emphasized how these models can benefit individuals from underrepresented groups in the software industry, including women, by providing increased access to work opportunities through the flexibility of remote work [21, 24]. In our study, women in software testing shared insights that highlighted how software testing often offered more entry-level job opportunities and allowed them to balance their professional commitments with house responsibilities, such as childcare. Therefore, to foster greater equity and diversity in the software industry, investigating strategies to support these women could be crucial.

Secondly, our findings emphasize the significance of raising awareness about software testing among university students [8, 26]. This issue has been previously identified in the literature, which has observed that software engineering students often have outdated perspectives on the role of software testing. In our study, we identified evidence suggesting that enhancing awareness of software testing among university students could be particularly compelling for girls. One participant pointed out that software testing is an area with fewer gender stereotypes, which might mean a more inclusive environment. However, it is essential to acknowledge that this evidence requires further exploration through additional studies.

Thirdly, it is important to explore why, in some instances, women are being pushed towards software testing. Our study underscores that women frequently choose a testing career voluntarily. However, it also highlights situations where women transition from roles in software development, like programming, to software testing—a field often viewed as unpopular or less prestigious [17]. Further investigation is essential to guarantee fair role assignments in software teams, promoting an equitable and inclusive environment. Understanding the dynamics behind these transitions will contribute to fostering equity and challenging stereotypes in the software engineering landscape.

## 5 LIMITATIONS AND THREATS TO VALIDITY

At this stage, it is important to acknowledge certain limitations in our study, particularly in terms of generalizing the results. Our sample size is limited, and we predominantly relied on convenience sampling, which prevents broad generalizations. However, we argue that our findings can act as a catalyst for discussions and provide valuable insights to researchers, particularly considering the lack of diversity in software engineering – exploring topics related to equity, diversity, and inclusion in the field requires several investigations and our study contributes to this ongoing scenario. Additionally, for practitioners, our findings can assist in initiating discussions within their contexts and obtaining additional evidence to create and implement practices supporting women in their teams.

## 6 CONCLUSION

In the context of the underrepresentation of women in the software industry and the dynamic landscape of software engineering, this study aimed to conduct a preliminary investigation on the motivations guiding women in pursuing careers in software testing. Our main goal was to identify the intricate factors shaping the experiences and decisions of women within software testing.

Our findings indicate that women are attracted to software testing for several reasons, including the abundance of entry-level job opportunities, the field's compatibility with work-life balance, recommendations from female peers, and the perceived decrease of gender stereotypes. Furthermore, our findings highlight that women remain in the software testing field due to the meaningful impact of their work on others, continuous learning opportunities, and the challenging aspects of the job, including involvement in testing automation (e.g., programming). However, the study also highlights areas that warrant improvement, particularly in fostering inclusiveness and supporting career development.

Although preliminary, the significance of our findings extends beyond the current study, providing valuable insights for discussions among researchers and practitioners. Understanding the diverse motivations of women in software testing is essential for creating inclusive workplaces and fostering the professional growth of women in the field. As the software industry continues to evolve, leveraging these insights can contribute to building a more diverse, inclusive, and equitable landscape in software testing and beyond.